\documentclass{article}
\begin{document}
{\large
\begin{center}
{\bf
Form factors of $S$-wave charmed baryon multiplet
$J^P=\frac{1}{2}^+$.}
\end{center}

\vskip3ex
\noindent
S.M. Gerasyuta, E.E. Matskevich
\vskip3ex
\noindent
Department of Theoretical Physics, St. Petersburg State University,
198904, St. Petersburg, Russia

\noindent
Department of Physics, LTA, 194021, St. Petersburg, Russia

\vskip4ex
\begin{center}
{\bf Abstract}
\end{center}
\vskip4ex
Electric form factors of $S$-wave charmed baryons are calculated within
the relativistic quark model in the region of low and intermediate
momentum transfers, $Q^2 \le 1 \, GeV^2$. The charge radii of low-lying
charmed baryons are determined.
\vskip2ex
\noindent
e-mail address: gerasyuta@SG6488.spb.edu

\noindent
e-mail address: matskev@pobox.spbu.ru
\vskip2ex
\noindent
\vskip3ex

The inclusion of relativistic effects in composite systems is fairly
important in considering the quark structure of hadrons [1 -- 10]. The
dynamical variables (form factors, scattering amplitudes) of composite
particles can be expressed in terms of Bethe-Salpeter functions or
quasipotentials. The form factors of composite particles were considered
by a number of authors, using, in particular, the ladder approximation
for Bethe-Salpeter equation [11] and ideas of conformal invariance [12].
A number of results were obtained in the framework of three-dimentional
formalisms [13]. Apparently, a fairly convenient way of describing
relativistic effects in composite systems may be the use of dispersion
integrals in the masses of composite particles. On the one hand, the
technique of dispersion integration is relativistically invariant and not
related to the consideration of any specific coordinate system. On the
other hand there are no problems whith the appearance of extra states,
since in the dispersion relations the contributions of intermediate states
are under control. The dispersion relation technique makes it possible to
determine form factors for composite particles [14].

In the paper [15] we have constructed a relativistic generalization of
the three-particles Faddeev equations in the form of dispersion relations
in the pair energy of the two interacting particles. By the method of
extraction of the leading singularities of the amplitude we have calculated
the mass spectrum of $S$-wave baryons, the multiplets $J^P=\frac{1}{2}^+$
and $J^P=\frac{3}{2}^+$ [15, 16], and we have obtained the electric form
factors of nucleons at low and intermediate momentum transfers [17, 18].

In the present paper we calculate the form factors and the charge radii of
$S$-wave charmed baryon multiplet $J^P=\frac{1}{2}^+$.

We consider electric form factor of the three-particle systems (charmed
baryons) shown in Fig. 1,a. We assume that the momentum of the baryon is
large ($P_z \to\infty$). The momenta $P=k_1+k_2+k_3$ and $P'=P+q$ correspond
to the initial and final momenta of the system, and $P^2=s$, $P'^2=s'$; $s$
and $s'$ are the initial and final energies of the system. We assume that
$P=(P_0, {\bf P}_{\bot}=0, P_z)$ and $P'=(P'_0, {\bf P}'_{\bot}, P'_z)$.

Then we have several conservation laws for the incoming momenta:

$${\bf k}_{1\bot}+{\bf k}_{2\bot}+{\bf k}_{3\bot}=0\, ,$$

$$P_z-k_{1z}-k_{2z}-k_{3z}=P_z (1-x_1-x_2-x_3)=0\, ,$$

$$P_0-k_{10}-k_{20}-k_{30}=P_z (1-x_1-x_2-x_3)+\frac{1}{2P_z}
\left( s-\frac{m^2_{1\bot}}{x_1}-\frac{m^2_{2\bot}}{x_2}
-\frac{m^2_{3\bot}}{x_3}\right)=0\, ,$$

$$m^2_{i\bot}=m^2_i +k^2_{i\bot}\, , \, x_i=\frac{k_{iz}}{P_z}\, , \,
i=1, 2, 3. \eqno(1)$$

Similarly, for the outgoing momenta,

$${\bf k}'_{1\bot}+{\bf k}_{2\bot}+{\bf k}_{3\bot}-{\bf q}_{\bot}=0\, ,$$

$$P'_z-k'_{1z}-k_{2z}-k_{3z}=P_z (z-x'_1-x_2-x_3)=0\, ,$$

$$x'_1=\frac{k'_{1z}}{P_z}\, , \, z=\frac{P'_z}{P_z}=
\frac{s'+s-q^2}{2s}\, ,$$

$$P'_0-k'_{10}-k_{20}-k_{30}=P_z (z-x'_1-x_2-x_3)+\frac{1}{2P_z}
\left( \frac{s'+{\bf q}^2_{\bot}}{z}-\frac{m'^2_{1\bot}}{x_1}-
\frac{m^2_{2\bot}}{x_2}-\frac{m^2_{3\bot}}{x_3}\right)=0\, ,$$

$$m'^2_{1\bot}=m^2_1 +k'^2_{1\bot}\, , \, {\bf q}_{\bot}=
{\bf P}'_{\bot} \, , s'=P'^2=P'^2_0-{\bf P}'^2_{\bot}-P'^2_z \, . \eqno(2)$$

Form factor for a system of three quarks can be obtained by means of
a double dispersion relation:

$$F(q^2)=\int\limits_{(m_1+m_2+m_3)^2}^{\Lambda_s} \frac{ds\, ds'}{4\pi^2}
\, \frac{disc_s \, disc_{s'}\, F(s, s', q^2)}{(s-M^2)(s'-M^2)}
\, , \eqno(3)$$

$$disc_s \, disc_{s'}\, F(s, s', q^2)=G(s, s_{12})G(s', s_{12})
\int d\rho(P, P', k_1, k_2)\, . \eqno(4)$$

The integral for the invariant phase-space has the following form
(Fig. 1,a):

$$\int d\rho(P, P', k_1, k_2)=\int (2\pi)^4 \, \delta^4(P-k_1-k_2-k_3)\,
\frac{d^3 k_1}{(2\pi)^3(2k_{10})}\,\frac{d^3 k_2}{(2\pi)^3(2k_{20})}\,
\frac{d^3 k_3}{(2\pi)^3(2k_{30})}$$

$$\times (2\pi)^4 \, \delta^4(P'-k'_1-k'_2-k'_3)\,
\frac{d^3 k'_1}{(2\pi)^3(2k'_{10})}\,
\frac{d^3 k'_2}{(2\pi)^3(2k'_{20})}\,\frac{d^3 k'_3}{(2\pi)^3(2k'_{30})}$$

$$\times (2k_{20})(2\pi)^3 \,\delta^3({\bf k}_2-{\bf k}'_2)\,
(2k_{30})(2\pi)^3 \,\delta^3({\bf k}_3-{\bf k}'_3)\, . \eqno(5)$$

In the presence of a spectator diquark (Fig. 1,b) we obtain:

$$\int d\rho(P, P', k_1)=\int\limits_{(m_2+m_3)^2}^{\Lambda_{23}}
ds_{23}\, \frac{1}{(2\pi)^2}\int (2\pi)^4 \, \delta^4(k_{23}-k_2-k_3)\,
\frac{d^3 k_2}{(2\pi)^3(2k_{20})}\,\frac{d^3 k_3}{(2\pi)^3(2k_{30})}$$

$$\times\int(2\pi)^4 \, \delta^4(P-k_1-k_{23})\,
\frac{d^3 k_1}{(2\pi)^3(2k_{10})}\,\frac{d^3 k_{23}}{(2\pi)^3(2k_{230})}$$

$$\times (2\pi)^4 \, \delta^4(P'-k'_1-k'_{23})\,
\frac{d^3 k'_1}{(2\pi)^3(2k'_{10})}\,
\frac{d^3 k'_{23}}{(2\pi)^3(2k'_{230})}\, (2k_{230})(2\pi)^3
\delta^3({\bf k}_{23}-{\bf k}'_{23})\, . \eqno(6)$$

Upon some transformations, we obtain the form factors of charmed baryons as:

$$F(q^2)=\frac{1}{8}\,\frac{1}{(2\pi)^6}(J_3+J_6)\, , \eqno(7)$$

\noindent
here the $J_3$ and $J_6$ are the contributions of Fig. 1,b and Fig. 1,a
respectively:

$$J_3=I_{23}\, \int \limits_{0}^{\Lambda_{k_{\bot}}} dk^2_{\bot}
\int \limits_{0}^{2\pi} d\varphi \int \limits_{0}^{1} \frac{dx}{x(1-x)}\,
\frac{b\lambda+1}{b+\lambda f}\, \frac{1}{(s-M^2)(s'-M^2)}$$

$$\times(G(s)G(s'))_3\,\Theta(\Lambda_s -s)\,\Theta(\Lambda_s -s')
\, , \eqno(8)$$

$$J_6=\int \limits_{0}^{\Lambda_{k_{\bot}}} dk^2_{1\bot}
\int \limits_{0}^{\Lambda_{k_{\bot}}} dk^2_{2\bot}
\int \limits_{0}^{2\pi} d\varphi_1 \int \limits_{0}^{2\pi} d\varphi_2
\int \limits_{0}^{1} dx_1 \int \limits_{0}^{1} dx_2\,
\frac{1}{x_1 x_2(1-x_1)(1-x_2)}\,
\frac{\tilde b \tilde \lambda+1}{\tilde b+\tilde \lambda \tilde f}$$

$$\times\frac{1}{(\tilde s-M^2)(\tilde s'-M^2)}\times(G(s)G(s'))_6\,
\Theta(\Lambda_s -\tilde s)\,\Theta(\Lambda_s -\tilde s')\, . \eqno(9)$$

The $I_{23}$ corresponds to the diquark phase space. We introduce the
following notations:

$$b=1+\frac{m_1^2-m_{23}^2}{s}\, ; \,
f=b^2-\frac{4k^2_{\bot} cos^2 \varphi}{s}\, ;
\lambda=\frac{-b+\sqrt{(b^2-f)(1-(s/q^2)f)}}{f}\, ;$$

$$s=\frac{m^2_{1\bot}+x(m_{23}^2-m_1^2)}{x(1-x)}\, ;
s'=s+q^2 (1+2\lambda)\, ; \eqno(10)$$

$$\tilde b=x_1+\frac{m_{1\bot}^2}{\tilde s x_1}\, ; \,
\tilde f=\tilde b^2-\frac{4k^2_{1\bot} cos^2 \varphi_1}{\tilde s}\, ;
\tilde \lambda=\frac{-\tilde b+\sqrt{(\tilde b^2-\tilde f)
(1-(\tilde s/q^2)\tilde f)}}{\tilde f}\, ;$$

$$\tilde s=\frac{m_{1\bot}^2}{x_1}+\frac{m_{2\bot}^2}{x_1}+
\frac{m_3^2+k_{1\bot}^2+k_{2\bot}^2+2(k_{1\bot}^2 k_{2\bot}^2)^{1/2}
cos(\varphi_2-\varphi_1)}{(1-x_1)(1-x_2)}\, ;
\tilde s'=\tilde s+q^2 (1+2\tilde \lambda)\, .$$

The quark masses and the two-body cutoff are similar to the paper [19]:
$m_{u, d}=0.495\, GeV$, $m_s=0.770\, GeV$, $m_c=1.655\, GeV$;
$\lambda=10.7$. The dimensional cutoff parameters for the pair energy of
nonstrange, strange and charmed diquarks are
$\Lambda_{ab}=\lambda \frac{(m_a+m_b)^2}{4}$:
$\Lambda_{uc}=12.4\, GeV^2$, $\Lambda_{sc}=15.7\, GeV^2$,
$\Lambda_{cc}=29.3\, GeV^2$. The transverse momentum cutoffs are
$\Lambda_{k_{\bot}}=1.04$ for the $\Sigma_c^{++}$, $\Sigma_c^{+}$,
$\Lambda_c^{+}$; for the $\Xi_c^{+A}$, $\Xi_c^{+S}$
$\Lambda_{k_{\bot}}=1.37$; for the $\Omega_{ccq}^{++}$, $\Omega_{ccq}^{+}$
$\Lambda_{k_{\bot}}=2.70$; for the $\Omega_{ccs}^{+}$
$\Lambda_{k_{\bot}}=2.47$.

The vertex functions $(G(s)G(s'))_3$ and $(G(s)G(s'))_6$ are determined
by the wave-functions of the corresponding charmed baryons (Appendix 1).
To find the form factor of the charmed baryons we must include the
interaction of each quark with an external electric field by means of the
form factor of nonstrange, strange and charmed quarks: for the
$u$, $d$-quarks $f_q (q^2)=exp(\alpha _q q^2)$, $\alpha _q=0.33\, GeV^2$,
for the $s$-quark $f_s (q^2)=exp(\alpha _s q^2)$, $\alpha _s=0.20\, GeV^2$,
and for the $c$-quark $f_á (q^2)=1$.

The equation (7) use for the a numerical calculation of the form factors
with the normalization $G^E_p (0)=1$. The behaviour of the electric form
factor of the $\Sigma_c^{++}$, $\Sigma_c^+$ is shown in Fig. 2. The
calculated value of the charge radii of the charmed baryons are found to
be (Table 1):

$$R_{\Omega_{ccs}^{+}}<R_{\Omega_{ccq}^{++}}
<R_{\Xi_c^{+A}}, R_{\Lambda_c^{+}}<R_{\Omega_{ccq}^{+}}, R_{\Xi_c^{+S}}
<R_{\Sigma_c^{++}}<R_{\Sigma_c^{+}}\, . \eqno(11)$$

The charmed radii of the neutral baryons are found to be practically
equal to zero.

It is worth noting that we do not use any new parameters here. All of the
parameters involved were borrowed from the calculations of the mass spectrum
of the $S$-wave charmed baryons [19].

\vskip2ex
{\bf Acknowledgment}
\vskip2ex

The authors thanks D.V. Ivanov for the assistance with the calculations.
The work was carried with the support of the Russion Ministry of Education
(grant 2.1.1.68.26).

\vskip3ex

{\Large
\bf References.}

}
\vskip3ex

\noindent
1. F. Gross, Phys. Rev. Lett. {\bf 140}, 410 (1965).

\noindent
2. H. Melosh, Phys. Rev. D{\bf 9}, 1095 (1974).

\noindent
3. G.B. West, Ann. Phys. (N.Y.) D{\bf 74}, 464 (1972).

\noindent
4. S.J. Brodsky and G.R. Farrar, Phys. Rev. D{\bf 11}, 1309 (1975).

\noindent
5. M.V. Terent'ev, Yad. Fiz. {\bf 24}, 207 (1976) [Sov. J. Nucl. Phys.
{\bf 24}, 106 (1976)].

\noindent
6. V.A. Karmanov, Zh. Eksp. Teor. Fiz. {\bf 71}, 399 (1976) [Sov. Phys.
JETP{\bf 44}, 210

(1976)].

\noindent
7. I.S. Aznauryan and L.N. Ter-Isaakyan, Yad. Fiz. {\bf 31}, 1680 (1980)
[Sov. J. Nucl.

Phys. {\bf 31}, 871 (1980)].

\noindent
8. A. Donnachie, R.R. Horgen, and P.V. Landshoft, Z. Phys. C{\bf 10}, 71
(1981).

\noindent
9. L.L. Frankfurt and M.I. Strikman, Phys. Rep. C{\bf 76}, 215 (1981).

\noindent
10. L.A. Kondratyuk and M.I. Strikman, Nucl. Phys. A{\bf 426}, 575 (1984).

\noindent
11. R.N. Faustov, Ann. Phys. {\bf 78}, 176 (1973).

\noindent
12. A.A. Migdal, Phys. Lett. B{\bf 37}, 98 (1971).

\noindent
13. R.N. Faustov, Teor. Mat. Fiz. {\bf 3}, 240 (1970).

\noindent
14. V.V. Anisovich and A.V. Sarantsev, Yad. Fiz. {\bf 45}, 1479 (1987)
[Sov. J. Nucl.

Phys. {\bf 45}, 918 (1987)].

\noindent
15. S.M. Gerasyuta, Yad. Fiz. {\bf 55}, 3030 (1992) [Sov. J. Nucl. Phys.
{\bf 55}, 1693 (1992)].

\noindent
16. S.M. Gerasyuta, Z. Phys. C{\bf 60},  683 (1993).

\noindent
17. S.M. Gerasyuta, Nuovo. Cim. A{\bf 106}, 37 (1993).

\noindent
18. S.M. Gerasyuta and D.V. Ivanov, Vest. St. Peterburg Univ. Ser. 4,
ü 2 (11), 3 (1996).

\noindent
19. S.M. Gerasyuta and D.V. Ivanov, Yad. Fiz. {\bf 62}, 1693 (1999).

\newpage

\vskip30pt
\begin{picture}(600,110)
\put(10,10){\line(1,0){60}}
\put(95,10){\oval(50,40)[t]}
\put(95,10){\oval(50,40)[b]}
\put(120,10){\line(1,0){60}}
\put(69,10){\line(1,3){26}}
\put(121,10){\line(-1,3){26}}
\put(40,10){\vector(1,0){6}}
\put(140,10){\vector(1,0){6}}
\put(90,30){\vector(1,0){6}}
\put(90,-10){\vector(1,0){6}}
\put(79,40){\vector(1,3){3}}
\put(107,51){\vector(1,-3){3}}
\put(96,90){\oval(6,4)[l]}
\put(94,94){\oval(6,4)[r]}
\put(96,98){\oval(6,4)[l]}
\put(94,102){\oval(6,4)[r]}
\put(180,10){\vector(1,0){6}}
\put(100,94){$q$}
\put(35,-6){$P$}
\put(140,-6){$P'$}
\put(66,40){$k_1$}
\put(114,40){$k'_1$}
\put(90,17){$k_2$}
\put(90,-23){$k_3$}
\put(170,-4){$z$}

\put(210,10){\line(1,0){170}}
\put(269,10){\line(1,3){26}}
\put(321,10){\line(-1,3){26}}
\put(240,10){\vector(1,0){6}}
\put(340,10){\vector(1,0){6}}
\put(279,40){\vector(1,3){3}}
\put(307,51){\vector(1,-3){3}}
\put(290,10){\vector(1,0){6}}
\put(296,90){\oval(6,4)[l]}
\put(294,94){\oval(6,4)[r]}
\put(296,98){\oval(6,4)[l]}
\put(294,102){\oval(6,4)[r]}
\put(380,10){\vector(1,0){6}}
\put(300,94){$q$}
\put(235,-6){$P$}
\put(340,-6){$P'$}
\put(266,40){$k_1$}
\put(314,40){$k'_1$}
\put(290,-6){$k_{23}$}
\put(370,-4){$z$}
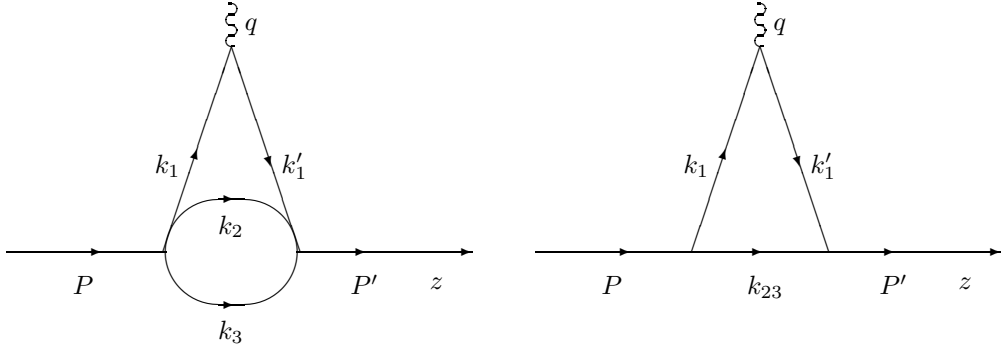
\put(10,-40){Fig. 1 a, b. Triangle diagrams the form factors
of charmed baryons.}
\end{picture}

\vskip100pt
\begin{picture}(600,140)
\put(10,10){\line(1,0){110}}
\put(10,10){\line(0,1){110}}
\put(7,110){\line(1,0){6}}
\put(8.5,100){\line(1,0){3}}
\put(8.5,90){\line(1,0){3}}
\put(8.5,80){\line(1,0){3}}
\put(8.5,70){\line(1,0){3}}
\put(7,60){\line(1,0){6}}
\put(8.5,50){\line(1,0){3}}
\put(8.5,40){\line(1,0){3}}
\put(8.5,30){\line(1,0){3}}
\put(8.5,20){\line(1,0){3}}
\put(110,7){\line(0,1){6}}
\put(100,8.5){\line(0,1){3}}
\put(90,8.5){\line(0,1){3}}
\put(80,8.5){\line(0,1){3}}
\put(70,8.5){\line(0,1){3}}
\put(60,7){\line(0,1){6}}
\put(50,8.5){\line(0,1){3}}
\put(40,8.5){\line(0,1){3}}
\put(30,8.5){\line(0,1){3}}
\put(20,8.5){\line(0,1){3}}
\put(10,110){\circle*{2}}
\put(20,104){\circle*{2}}
\put(30,98){\circle*{2}}
\put(40,94){\circle*{2}}
\put(50,90){\circle*{2}}
\put(60,87){\circle*{2}}
\put(70,82){\circle*{2}}
\put(80,79){\circle*{2}}
\put(90,76){\circle*{2}}
\put(100,74){\circle*{2}}
\put(110,71){\circle*{2}}
\put(-6,-2){0.0}
\put(-8,55){0.5}
\put(-8,105){1.0}
\put(55,-4){0.5}
\put(105,-4){1.0}
\put(-10,130){$F(Q^2)$}
\put(130,-4){$Q^2$, $GeV^2$}
\put(60,100){$\Sigma_c^{++}$}
\put(240,10){\line(1,0){110}}
\put(240,10){\line(0,1){110}}
\put(237,110){\line(1,0){6}}
\put(238.5,100){\line(1,0){3}}
\put(238.5,90){\line(1,0){3}}
\put(238.5,80){\line(1,0){3}}
\put(238.5,70){\line(1,0){3}}
\put(237,60){\line(1,0){6}}
\put(238.5,50){\line(1,0){3}}
\put(238.5,40){\line(1,0){3}}
\put(238.5,30){\line(1,0){3}}
\put(238.5,20){\line(1,0){3}}
\put(340,7){\line(0,1){6}}
\put(330,8.5){\line(0,1){3}}
\put(320,8.5){\line(0,1){3}}
\put(310,8.5){\line(0,1){3}}
\put(300,8.5){\line(0,1){3}}
\put(290,7){\line(0,1){6}}
\put(280,8.5){\line(0,1){3}}
\put(270,8.5){\line(0,1){3}}
\put(260,8.5){\line(0,1){3}}
\put(250,8.5){\line(0,1){3}}
\put(240,110){\circle*{2}}
\put(250,103){\circle*{2}}
\put(260,97.7){\circle*{2}}
\put(270,93){\circle*{2}}
\put(280,90.7){\circle*{2}}
\put(290,87.2){\circle*{2}}
\put(300,85.4){\circle*{2}}
\put(310,83.7){\circle*{2}}
\put(320,81.9){\circle*{2}}
\put(330,80.2){\circle*{2}}
\put(340,78.4){\circle*{2}}
\put(224,-2){0.0}
\put(222,55){0.5}
\put(222,105){1.0}
\put(285,-4){0.5}
\put(335,-4){1.0}
\put(220,130){$F(Q^2)$}
\put(360,-4){$Q^2$, $GeV^2$}
\put(290,100){$\Sigma_c^{+}$}
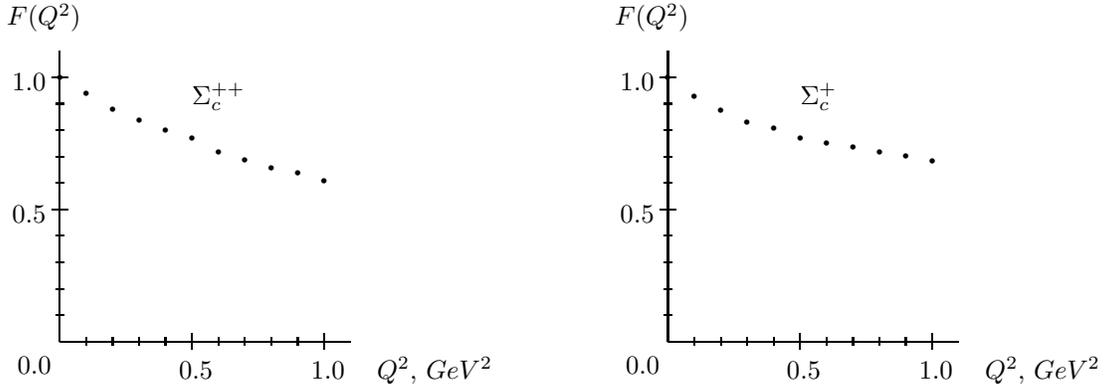
\put(10,-40){Fig. 2. Electric form factors of the charmed baryons
$\Sigma_c^{++}$, $\Sigma_c^{+}$ at small and}
\put(10,-60){intermediate transfers
$Q^2 < 1 \, GeV^2 \, (Q^2 \equiv -q^2)$.}
\end{picture}

\vskip120pt

{Table 1. The $S$-wave charmed baryons charge radii
$J^P=\frac{1}{2}^+$.
\vskip20pt

\begin{tabular}{|c|c|c|}
\hline
particle & mass (GeV) & charge radius (fm) \\
\hline
$\Lambda_c^+$ & 2.284 & 0.34 \\
\hline
$\Sigma_c^{++}$ & 2.458 & 0.39 \\
\hline
$\Sigma_c^+$ & 2.458 & 0.41 \\
\hline
$\Xi_c^{+A}$ & 2.467 & 0.34 \\
\hline
$\Xi_c^{+S}$ & 2.565 & 0.35 \\
\hline
$\Omega_{ccq}^{++}$ & 3.527 & 0.32 \\
\hline
$\Omega_{ccq}^+$ & 3.527 & 0.35 \\
\hline
$\Omega_{ccs}^+$ & 3.598 & 0.25 \\
\hline
\end{tabular}

\newpage
\noindent
{\bf Appendix 1. The vertex functions of the charmed baryon multiplet
$J^p=\frac{1}{2}^+$.}

\vskip5ex

$\Sigma_c^{++}$:

$$(G(s)G(s'))_3=\frac{1}{18}\, \frac{1}{4}\, \left(
A_1^{(c)^2} (s, s_0)\, f_q(q^2)\, e_u\, 12+
A_0^{(c)^2} (s, s_0)\, f_q(q^2)\, e_u\, 36\right.$$

$$\left.+A_1^2 (s, s_0)\, f_c(q^2)\, e_c\, 24
\right)$$

$$(G(s)G(s'))_6=\frac{1}{18}\, \frac{1}{4}\, \left(
A_1^{(c)^2} (s, s_0) (f_q(q^2)\, e_u\, 12+f_c(q^2)\, e_c\, 12)\right.$$

$$\left.+A_0^{(c)^2} (s, s_0) (f_q(q^2)\, e_u\, 36+f_c(q^2)\, e_c\, 36)
+A_1^2 (s, s_0)\, f_q(q^2)\, e_u\, 48
\right)$$

\vskip5ex

$\Sigma_c^+$:

$$(G(s)G(s'))_3=\frac{1}{36}\, \frac{1}{4}\, \left(
A_1^{(c)^2} (s, s_0)\, f_q(q^2) (e_u\, 12+e_d\, 12)\right.$$

$$\left.+A_0^{(c)^2} (s, s_0)\, f_q(q^2) (e_u\, 36+e_d\, 36)
+A_1^2 (s, s_0)\, f_c(q^2)\, e_c\, 48
\right)$$

$$(G(s)G(s'))_6=\frac{1}{36}\, \frac{1}{4}\, \left(
A_1^{(c)^2} (s, s_0) (f_q(q^2) (e_u\, 12+e_d\, 12)+
f_c(q^2)\, e_c\, 24)\right.$$

$$\left.+A_0^{(c)^2} (s, s_0) (f_q(q^2) (e_u\, 36+e_d\, 36)
+f_c(q^2)\, e_c\, 72)
+A_1^2 (s, s_0)\, f_q(q^2) (e_u\, 48+e_d\, 48)
\right)$$

\vskip3ex

For the $\Sigma_c^{++,+}$: $A_1 (s, s_0)=0.431$, $A_1^{(c)} (s, s_0)=2.203$,
$A_0^{(c)} (s, s_0)=2.534$.

\vskip5ex

$\Lambda_c^+$:

$$(G(s)G(s'))_3=\frac{1}{12}\, \frac{1}{4}\, \left(
A_1^{(c)^2} (s, s_0) (f_q(q^2) (e_u\, 6+e_d\, 6)+
f_c(q^2)\, e_c\, 12)\right.$$

$$\left.+A_0^{(c)^2} (s, s_0) (f_q(q^2) (e_u\, 2+e_d\, 2)
+f_c(q^2)\, e_c\, 4)
+A_0^2 (s, s_0)\, f_q(q^2) (e_u\, 8+e_d\, 8)
\right)$$

$$(G(s)G(s'))_6=\frac{1}{12}\, \frac{1}{4}\, \left(
A_1^{(c)^2} (s, s_0) (f_q(q^2) (e_u\, 18+e_d\, 18)+
f_c(q^2)\, e_c\, 12)+A_0^{(c)^2} (s, s_0)\right.$$

$$\left.
\cdot (f_q(q^2) (e_u\, 6+e_d\, 6)
+f_c(q^2)\, e_c\, 4)
+A_0^2 (s, s_0) (f_q(q^2) (e_u\, 8+e_d\, 8)+f_c(q^2)\, e_c\, 16)
\right)$$

\vskip3ex

For the $\Lambda_c^+$: $A_0 (s, s_0)=0.984$, $A_1^{(c)} (s, s_0)=2.23$,
$A_0^{(c)} (s, s_0)=2.548$.

\newpage

$\Xi_c^{+A}$:

$$(G(s)G(s'))_3=\frac{1}{12}\, \frac{1}{4}\, \left(
A_1^{(c)^2} (s, s_0) (f_q(q^2)\,e_u\, 6+f_c(q^2)\,e_c\, 6)\right.$$

$$\left. +A_1^{(sc)^2} (s, s_0) (f_s(q^2)\,e_s\, 6+f_c(q^2)\,e_c\, 6)
+A_0^{(c)^2} (s, s_0) (f_q(q^2)\,e_u\, 2+f_c(q^2)\,e_c\, 2)\right.$$

$$\left. +A_0^{(sc)^2} (s, s_0) (f_s(q^2)\,e_s\, 2+f_c(q^2)\,e_c\, 2)
+A_0^{(s)^2} (s, s_0) (f_q(q^2)\,e_u\, 8+f_s(q^2)\,e_s\, 8)
\right)$$

$$(G(s)G(s'))_6=\frac{1}{12}\, \frac{1}{4}\, \left(
A_1^{(c)^2} (s, s_0) (f_q(q^2)\,e_u\, 6+f_s(q^2)\,e_s\, 12
+f_c(q^2)\,e_c\, 6)\right.$$

$$\left. +A_1^{(sc)^2} (s, s_0) (f_q(q^2)\,e_u\, 12+f_s(q^2)\,e_s\, 6
+f_c(q^2)\,e_c\, 6) \right.$$

$$\left. +A_0^{(c)^2} (s, s_0) (f_q(q^2)\,e_u\, 2+f_s(q^2)\,e_s\, 4
+f_c(q^2)\,e_c\, 2) \right.$$

$$\left. +A_0^{(sc)^2} (s, s_0) (f_q(q^2)\,e_u\, 4+f_s(q^2)\,e_s\, 2
+f_c(q^2)\,e_c\, 2) \right.$$

$$\left. +A_0^{(s)^2} (s, s_0) (f_q(q^2)\,e_u\, 8+f_s(q^2)\,e_s\, 8
+f_c(q^2)\,e_c\, 16) \right)$$

\vskip3ex

For the $\Xi_c^{+A}$: $A_0^{(s)} (s, s_0)=1.032$, $A_1^{(c)} (s, s_0)=1.80$,
$A_0^{(c)} (s, s_0)=1.888$, $A_1^{(sc)} (s, s_0)=0.953$,
$A_0^{(sc)} (s, s_0)=1.147$.

\vskip5ex

$\Xi_c^{+S}$:

$$(G(s)G(s'))_3=\frac{1}{36}\, \frac{1}{4}\, \left(
A_1^{(c)^2} (s, s_0)\, f_s(q^2)\,e_s\, 12+
A_1^{(sc)^2} (s, s_0)\, f_q(q^2)\,e_u\, 12\right.$$

$$\left. +A_0^{(c)^2} (s, s_0)\, f_s(q^2)\,e_s\, 36+
A_0^{(sc)^2} (s, s_0)\, f_q(q^2)\,e_u\, 36+
A_1^2 (s, s_0)\, f_c(q^2)\,e_c\, 48\right)$$

$$(G(s)G(s'))_6=\frac{1}{36}\, \frac{1}{4}\, \left(
A_1^{(c)^2} (s, s_0) (f_q(q^2)\,e_u\, 12+f_c(q^2)\,e_c\, 12)
\right.$$

$$\left. +A_1^{(sc)^2} (s, s_0) (f_s(q^2)\,e_s\, 12+f_c(q^2)\,e_c\, 12)+
A_0^{(c)^2} (s, s_0) (f_q(q^2)\,e_u\, 36+f_c(q^2)\,e_c\, 36)
\right.$$

$$\left. +A_0^{(sc)^2} (s, s_0) (f_s(q^2)\,e_s\, 36+f_c(q^2)\,e_c\, 36)+
A_1^{(s)^2} (s, s_0) (f_q(q^2)\,e_u\, 48+f_s(q^2)\,e_s\, 48)
\right)$$

\vskip3ex

For the $\Xi_c^{+S}$: $A_1^{(s)} (s, s_0)=-0.373$,
$A_1^{(c)} (s, s_0)=1.281$, $A_0^{(c)} (s, s_0)=-1.173$,
$A_1^{(sc)} (s, s_0)=-2.954$, $A_0^{(sc)} (s, s_0)=1.042$.

\newpage

$\Omega_{ccq}^{++}$:

$$(G(s)G(s'))_3=\frac{1}{36}\, \frac{1}{4}\, \left(
A_1^{(c)^2} (s, s_0)\, f_c(q^2)\,e_c\, 24\right.$$

$$\left. +A_0^{(c)^2} (s, s_0)\, f_c(q^2)\,e_c\, 72
+A_1^{(cc)^2} (s, s_0)\, f_q(q^2)\,e_u\, 48\right)$$

$$(G(s)G(s'))_6=\frac{1}{36}\, \frac{1}{4}\, \left(
A_1^{(c)^2} (s, s_0) (f_c(q^2)\,e_c\, 24+f_q(q^2)\,e_u\, 24)\right.$$

$$\left. +A_0^{(c)^2} (s, s_0) (f_c(q^2)\,e_c\, 72+f_q(q^2)\,e_u\, 72)
+A_1^{(cc)^2} (s, s_0)\, f_c(q^2)\,e_c\, 96 \right)$$

\vskip3ex

For the $\Omega_{ccq}^+$: $e_u \to e_d$.

For the $\Omega_{ccq}^{++,+}$: $A_1^{(cc)} (s, s_0)=0.478$,
$A_1^{(c)} (s, s_0)=2.116$, $A_0^{(c)} (s, s_0)=2.433$.

\vskip5ex

$\Omega_{ccs}^+$:

$$(G(s)G(s'))_3=\frac{1}{36}\, \frac{1}{4}\, \left(
A_1^{(sc)^2} (s, s_0)\, f_c(q^2)\,e_c\, 24\right.$$

$$\left. +A_0^{(sc)^2} (s, s_0)\, f_c(q^2)\,e_c\, 72
+A_1^{(cc)^2} (s, s_0)\, f_s(q^2)\,e_s\, 48\right)$$

$$(G(s)G(s'))_6=\frac{1}{36}\, \frac{1}{4}\, \left(
A_1^{(sc)^2} (s, s_0) (f_c(q^2)\,e_c\, 24+f_s(q^2)\,e_s\, 24)\right.$$

$$\left. +A_0^{(sc)^2} (s, s_0) (f_c(q^2)\,e_c\, 72+f_s(q^2)\,e_s\, 72)
+A_1^{(cc)^2} (s, s_0)\, f_c(q^2)\,e_c\, 96 \right)$$

\vskip3ex

For the $\Omega_{ccs}^+$: $A_1^{(cc)} (s, s_0)=0.51$,
$A_1^{(sc)} (s, s_0)=1.686$, $A_0^{(sc)} (s, s_0)=1.95$.

}

\end{document}